\documentclass{ws-mpla}
\usepackage[super]{cite}
\usepackage{graphicx}
\usepackage{mathtools}
\begin{document}

\markboth{P.K. Sahoo, S.K. Tripathy, Parbati Sahoo}
{A periodic varying deceleration parameter in $f(R,T)$ gravity}

\catchline{}{}{}{}{}

\title{A periodic varying deceleration parameter in $f(R,T)$ gravity}

\author{\footnotesize P.K. Sahoo}
\address{Department of Mathematics,\\ Birla Institute of
Technology and Science-Pilani, \\ Hyderabad Campus, Hyderabad-500078,
India\\
pksahoo@hyderabad.bits-pilani.ac.in}

\author{S.K. Tripathy}

\address{Department of Physics,\\ Indira Gandhi Institute of Technology,\\ Sarang, Dhenkanal, Odisha-759146, India\\
tripathy \_\ sunil@rediffmail.com}

\author{Parbati Sahoo}

\address{Department of Mathematics,\\ Birla Institute of
Technology and Science-Pilani, \\ Hyderabad Campus, Hyderabad-500078,
India\\
sahooparbati1990@gmail.com}

\maketitle

\pub{Received (27 July 2018), Revised (29 August 2018)}{Accepted (02 September 2018)}

\begin{abstract}
The phenomenon of accelerated expansion of the present universe and a cosmic transit aspect  is explored in the framework of a modified gravity theory known as $f(R,T)$ gravity (where $R$ is the Ricci scalar and $T$ is the trace of the energy momentum tensor of the matter content). The cosmic transit phenomenon signifies a signature flipping behaviour of the deceleration parameter. We employ a periodic varying deceleration parameter and obtained the exact solution of field equations. The dynamical features of the model including the oscillatory behaviour of the EOS parameter are studied. We have also explored the obvious violation of energy momentum conservation in $f(R,T)$ gravity. The periodic behaviour of energy conditions for the model are also discussed with a wide range of the free parameters.

\end{abstract}

\ccode{PACS Nos.: 04.50.kd}

\section{Introduction}\label{I}

The phenomenon of accelerated expansion of the present universe becomes a centre of attention for all astrophysicists, cosmologists, and astronomers. Observations from Supernovae Ia, CMB, BAO and other astrophysical measurements have confirmed this fact.  The late time behavior of the universe is one of the major challenges in modern cosmology. In order to get a proper theoretical explanation of this phenomenon, various concepts and theories have been proposed in recent times by many researchers. Einstein's General theory of Relativity (GR) fails to explain this late time behaviour of the universe.  Modified gravity theories emerge as  alternatives to the conventional cosmology and  increasingly become popular to describe the late time cosmic speed up process. Geometrically modified theories are the generalizations of GR, in which the Einstein-Hilbert action is modified by replacing the Ricci scalar $R$ by a more general function, may be of Ricci scalar or of any other function with matter-geometry coupling. In particular, $f(R)$ gravity ($R$ is Ricci scalar) \cite{carroll04, nojiri07, bertolami07}, $f(\mathcal{T})$ gravity ($\mathcal{T}$ is torsion scalar) \cite{bengocheu09,linder10}, $f(G)$ gravity ($G$ is Gauss-Bonnet scalar ) \cite{bamba10a, bamba10b, rodrigues14}  and $f(R,T)$ gravity ($R$ is Ricci scalar and $T$ is trace of stress energy momentum tensor) \cite{Harko2011} are some widely used alternative modified theories. Among these geometrically modified theories, $f(R,T)$ theory has attracted a lot of attention of many cosmologists and  astrophysicists in recent times because of its ability to explain several issues in cosmology and astrophysics \cite{Zubair16, Singh016, Sahoo16, Sahoo16a, Sahu17, Mishra16, Moraes17, Yousaf17}. The big rip singularity of LRS Bianchi type-I space time has been obtained in this theory \cite{Sahoo/2015} . Also, it has an elegant geometrical structure and can be reduced to the GR and $f(R)$ theory under suitable functional forms of the functional $f(R,T)$. Moreover, the $T-$ dependence of the geometrical action in $f(R,T)$ theory may be due to the existence of some imperfect fluids and intrinsically may have some quantum effects like particle production \cite{Harko14}.

The matter energy coupling in $f(R,T)$ gravity plays a significant role to provide a complete theoretical description for the late time acceleration of the universe, without resorting to the existence of dark energy. In literature, there have been a lot of investigations on different aspects of $f(R,T)$ gravity such as scalar perturbation \cite{Alvarenga13}, energy conditions \cite{Sharif13, Kiani14, Alvarenga013}, thermodynamics \cite{Sharif12, Jamil012}, wormhole solutions \cite{Zubair/2016, Moraes17a, Moraes/2017}, and solutions in higher dimensions \cite{Yousaf16, Moraes15}. In this theory, the interactions of matter with space time curvature become a well motivation to consider cosmological consequence with different matter components \cite{Sahoo17, Sahoo017}. The role of violation of energy momentum conservation in modified theories have not yet been studied properly in literature.  Josset et al. \cite{Josset17}have shown that a violation of energy momentum conservation leads to accelerated expansion in modified gravity models. It will be more interesting to consider this phenomena in the context of $f(R,T)$ gravity. 

On the basis of the isotropic and spatially homogeneous universe, the  Friedmann Robertson Walker (FRW) metric is adequate for describing the present state of the universe. Due to this FRW models are globally acceptable with the perfect fluid matter on account of flat space-time. According to the literature, FRW cosmological model has been investigated in $f(R,T)$ gravity along with perfect fluid matter and linearly varying deceleration parameter \cite{Ramesh16}, and with magnetized strange quark matter and $\Lambda$ \cite{Aygun16}. After that, a time periodic varying deceleration parameter (PVDP) has been introduced by Shen and Zhao \cite{Shen2014} in order to account for an oscillating cosmological model with quintom matter. Oscillating cosmological models have been studied in literature \cite{Dodelson00, Nojiri06}. These models are very natural to resolve the coincidence problem due to periods of acceleration. 

We have organized the paper as follows. In Section-\ref{II}, we formulate the gravitational field equations of $f(R,T)$ theory for a flat universe. In Section-\ref{III}, solutions to the field equations are obtained by using the PVDP ansatz. Section-\ref{IV} contains a discussion on the dynamical properties of the model based on PVDP. We discuss briefly on the non-conservation of energy-momentum at the back drop of our present model in Section-\ref{V}. The energy conditions are discussed in Section-\ref{VI}. We discuss the stability of the model though linear homogeneous perturbation in Section-\ref{VIIa}. The results of the present work are summarized in the last section.

\section{Basic Field equations}\label{II}

In $f(R, T)$ gravity theory, we have a geometrically modified action 
\begin{equation}
\mathbb{S}=\frac{1}{2\kappa}\int f(R,T)\sqrt{-g}d^{4}x +\int \mathcal{L}_{m}\sqrt{-g}d^{4}x,\label{eq:1}
\end{equation}
which can be varied with respect to the metric tensor $g_{ij}$ to obtain the gravitational field equation for $f(R,T)$ gravity as
\begin{multline}
F(R,T)R_{ij}-\frac{1}{2}f(R,T)g_{ij}+(g_{ij}\Box -\nabla _{i}\nabla
_{j})F(R,T) \\ = \kappa T_{ij}-\mathcal{F}(R,T)T_{ij}- \mathcal{F}(R,T)\Theta _{ij}, \label{eq:2}
\end{multline}
where, $F(R,T)=\frac{\partial f(R,T)}{\partial R}$, $\mathcal{F}(R,T)=\frac{%
\partial f(R,T)}{\partial T}$, 
$\Box \equiv \nabla ^{i}\nabla _{i}$; $\nabla _{i}$ is the co-variant derivative. $\kappa=\frac{8\pi G}{c^4}$, where $G$ and $c$ are the Newtonian Gravitational constant and speed of light in vacuum respectively. The energy-momentum tensor for a perfect fluid distribution of the universe, $T_{ij}=-pg_{ij}+(\rho+p)u_iu_j$ and  $\Theta_{ij} =g^{\alpha \beta} \frac{\delta T_{\alpha \beta}}{\delta g^{ij}}$ are derived from the matter Lagrangian $\mathcal{L}_m$.  Following Harko et al. \cite{Harko2011}, we choose the matter Lagrangian as $\mathcal{L}_m=-p$ which yields $\Theta_{ij}=-pg_{ij}-2T_{ij}$. Here, $\rho$ and $p$ are the energy density and pressure respectively. $u^i$ is the four-velocity of the fluid satisfying $u_iu^i=1$ in comoving coordinates.

The functional $f(R,T)$ can be chosen in many ways corresponding to viable models. In the present work, we have considered the functional as $f(R,T)=R+2f(T)$ where $f(T)$ is an arbitrary function of the trace of the energy momentum tensor. The corresponding field equations become
\begin{equation}
R_{ij}-\frac{1}{2}Rg_{ij}=\kappa T_{ij}+2f_T T_{ij}+\left[f(T)+2pf_T\right]g_{ij} \label{eq:3}
\end{equation}
where $f_T$ denotes the partial derivative of $f$ with respect to $T$. Assuming $f(T)=\lambda T$, $\lambda$ being a constant, the field equations for a flat  Friedmann$-$Robertson$-$Walker (FRW) metric 
\begin{equation}
ds^2=dt^2-a^2(t)\left[dx^2+dy^2+dz^2\right] \label{eq:4}
\end{equation} 
are obtained as 

\begin{equation}
3H^2=(1+3\lambda)\rho-\lambda p, \label{eq:5}
\end{equation}
\begin{equation}
2\dot{H}+3H^2=\lambda \rho-(1+3\lambda)p.\label{eq:6}
\end{equation}
In the above equations, $a=a(t)$ is the scale factor of the universe and the Hubble parameter is defined through $H=\frac{\dot{a}}{a}$. We have chosen the unit system such that $\kappa=1$.  

\section{Periodically varying Deceleration parameter}\label{III}

Deceleration parameter (DP) is one of the geometrical parameters through which the dynamics of the universe can be quantified. It is defined as $q = -1-\frac{\dot{H}}{H^2}$ where the overhead dots denote time derivatives. In the context of the late time cosmic speed up phenomena with a cosmic transit from a phase of deceleration to acceleration at some redshift $z_{da} \sim 1$, one can speculate a signature flipping of the deceleration parameter. Obviously, at a decelerated phase, $q$ is positive and at the accelerating phase, it becomes negative. 
Geometrical parameters such as the deceleration parameter and jerk parameter are usually extracted from observations of high $z$ supernova. However, the exact time dependence of these parameters are not known to a satisfactory extent. In the absence of any explicit form of these parameters, many authors have used parametrized forms especially that of the DP to address different cosmological issues. Many parametrized forms of DP such as constant DP, linearly varying DP, quadratic varying DP etc are available in literature (for details one can refer to Ref.\cite{Pacif/2017}). Berman \cite{Berman83}, Berman and Gomide \cite{Berman88}, Pacif and Mishra \cite{Pacif/2015} proposed a special law of variation of Hubble parameter in FLRW-space time, which yields a constant form of DP. This law of variation for Hubble's parameter is valid for slowly varying DP models (for example; \cite{Akarsu10, Bishi/2017}). Linear parametrization of the DP shows quite natural phenomena toward the future evolution of the universe, either it expands forever or ends up with Big rip in finite future. Such a parametrisation has been  used frequently in literature \cite{Akarsu12,Sahoo/2015} .  
It is to mention here that the general dynamical behaviour can be assessed through the values of the deceleration parameter in the negative domain. While de Sitter expansion occurs for $q=-1$, accelerating power-law expansion can be achieved for $-1<q<0$. A super-exponential expansion of the universe occurs for $q<-1$.  Even though, there is uncertainty in the determination of the deceleration parameter from observational data, most of the studies in recent times constrain this parameter in the range $-0.8 \leq q \leq -0.4$. Keeping in view the signature flipping nature of $q$, in the present work, we assume a periodic time varying deceleration parameter \cite{Shen2014} 
\begin{equation}
q = m\cos kt-1 \label{eq:7}
\end{equation}
where $m$ and $k$ are positive constants.  Here $k$ decides the periodicity of the PVDP and can be considered as a cosmic frequency parameter. $m$ is an enhancement factor that enhances the peak of the PVDP. This model simulates a positive deceleration parameter $q=m-1$  (for $m>1$) at an initial epoch and evolves into a negative peak of $q=-m-1$. After the negative peak, it again increases and comes back to the initial states. The evolutionary behaviour of $q$ is periodically repeated. In other words,  the universe in the model, starts with a decelerating phase and evolves into a phase of super-exponential expansion in a cyclic history.

Integration of Eq. \eqref{eq:7} yields
\begin{equation}
H=\frac{k}{m\sin kt+k_1},\label{eq:8}
\end{equation}
where $k_1$ is a constant of integration. Here we have used the definitions $q= -1-\frac{\dot{H}}{H^2}$ and $\dot{a}=aH$ so that $\dot{H}=-mH^2\cos kt$. Without loss of generality, we may consider $k_1=0$ and the Hubble function becomes 
\begin{equation}
H=\frac{k}{m\sin kt}.\label{eq:9}
\end{equation}
The scale factor $a$ is obtained by integrating the Hubble function in Eq. \eqref{eq:9} as
\begin{equation}
a=a_0\left[\tan \left(\frac{1}{2}kt\right)\right]^{\frac{1}{m}},\label{eq:10}
\end{equation}
where $a_0$ is the scale factor at the present epoch  and can be taken as 1. Inverting Eq. \eqref{eq:10}, we obtain
\begin{equation}
t=\frac{2 \tan ^{-1}\left[\frac{1}{(z+1)^m}\right]}{k}.\label{eq:11}
\end{equation}

\begin{figure}
\centering
\includegraphics[width=75mm]{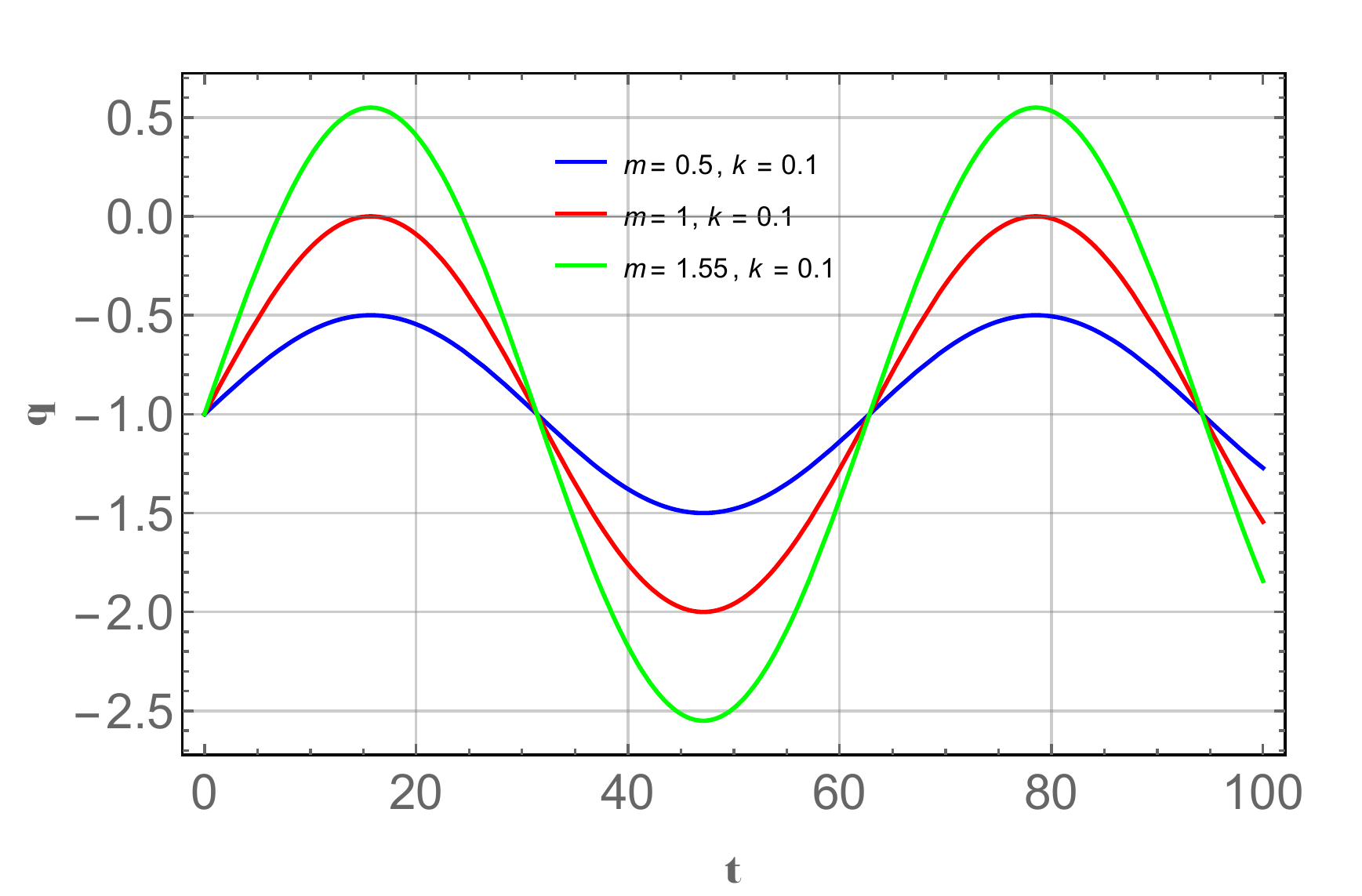}
  \caption{Evolution of deceleration parameter for three representative values of $m$ and $k=0.1$.}\label{fig1}
\end{figure}
\begin{figure}
\centering
  \includegraphics[width=75mm]{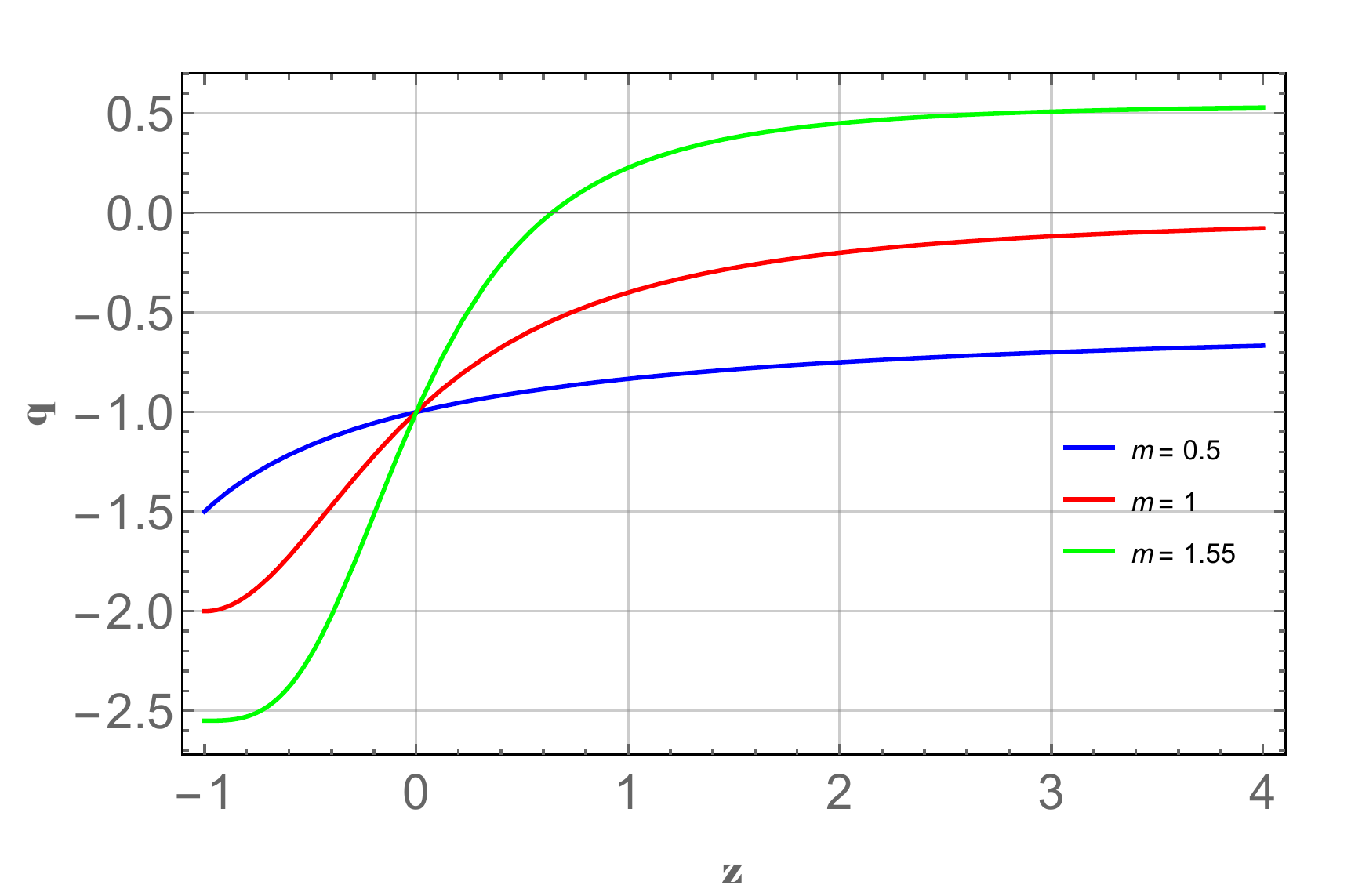}
  \caption{Evolution of deceleration parameter as a function of redshift. The cosmic transit behaviour is obtained for $m>1$.}\label{fig2}
\end{figure}


In the above equation, redshift is defined  through the relation $z=\frac{1}{a}-1$. An equivalent present epoch can be derived from Eq.\eqref{eq:11} as $t=\left(\frac{8n+1}{k}\right)\frac{\pi}{2}$ corresponding to $a_0=1$, where $n=0,1,2,3,\cdots$ is a positive integer including zero. It can be straightforward to express the deceleration parameter in Eq. \eqref{eq:7} in terms of redshift using Eq.\eqref{eq:11}. In Fig.\ref{fig1}, we have shown the evolutionary aspect of the deceleration parameter as a function of cosmic time for three different domain of the parameter $m$ namely $m<1, m=1$ and $m>1$. The periodic nature of the PVDP is clearly depicted in the figure. In Fig. \ref{fig2}, the evolutionary aspect of the deceleration parameter as a function of redshift is shown. The evolutionary behaviour of the PVDP is greatly affected by the choice of the parameter $m$.  In general, the deceleration parameter oscillates in between $m-1$ and $-m-1$. For $m=0$, deceleration parameter becomes a constant quantity with a value of $-1$ and can lead to a de Sitter kind of expansion. For $0<m\leq 1$, it varies periodically in the negative domain and provides accelerated models. However, for $m>1$, $q$ evolves from a positive region to a negative region showing a signature flipping at some redshift $z_{da}$. It is worth to mention here that, the transition redshift depends on the choice of  the parameter $m$. This parameter can be constrained from the cosmic transit behaviour and transit redshift $z_{da}$. We have adjusted the values of $m$ so as to get a $z_{da}$ compatible with that extracted from observations \cite{Capozziello14, Capozziello15, Farooq17, Moraes16}. In Fig.\ref{fig2}, the signature flipping of the deceleration parameter is shown  to occur at $z_{da}=0.64$ for $m=1.55$.  In the event of non availability of any observational data regarding cosmic oscillation and corresponding frequency, we consider $k$ as a free parameter. In the present work, we are interested for a time varying deceleration parameter that oscillates in between the decelerating and accelerating phase to simulate the cosmic transit phenomenon.  In order to assess the dynamical features of the model through numerical plots, we assume a small value for $k$, say $k=0.1$.
\begin{figure}
\centering
\includegraphics[width=75mm]{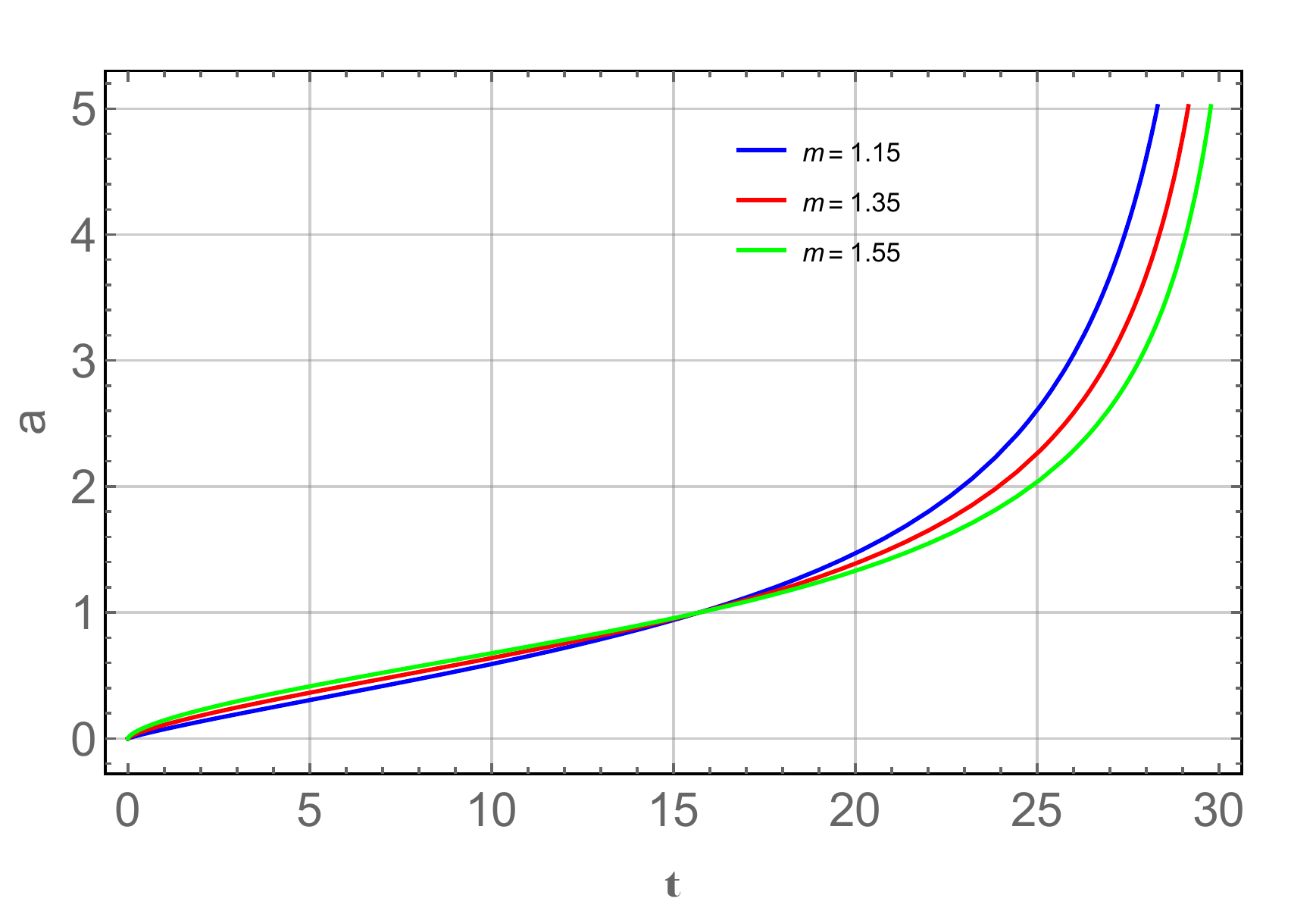}
  \caption{Scale factor as a function of cosmic time for $k=0.1$ and three representative values of $m$.}\label{fig3}
\end{figure}
\begin{figure}
\centering
  \includegraphics[width=75mm]{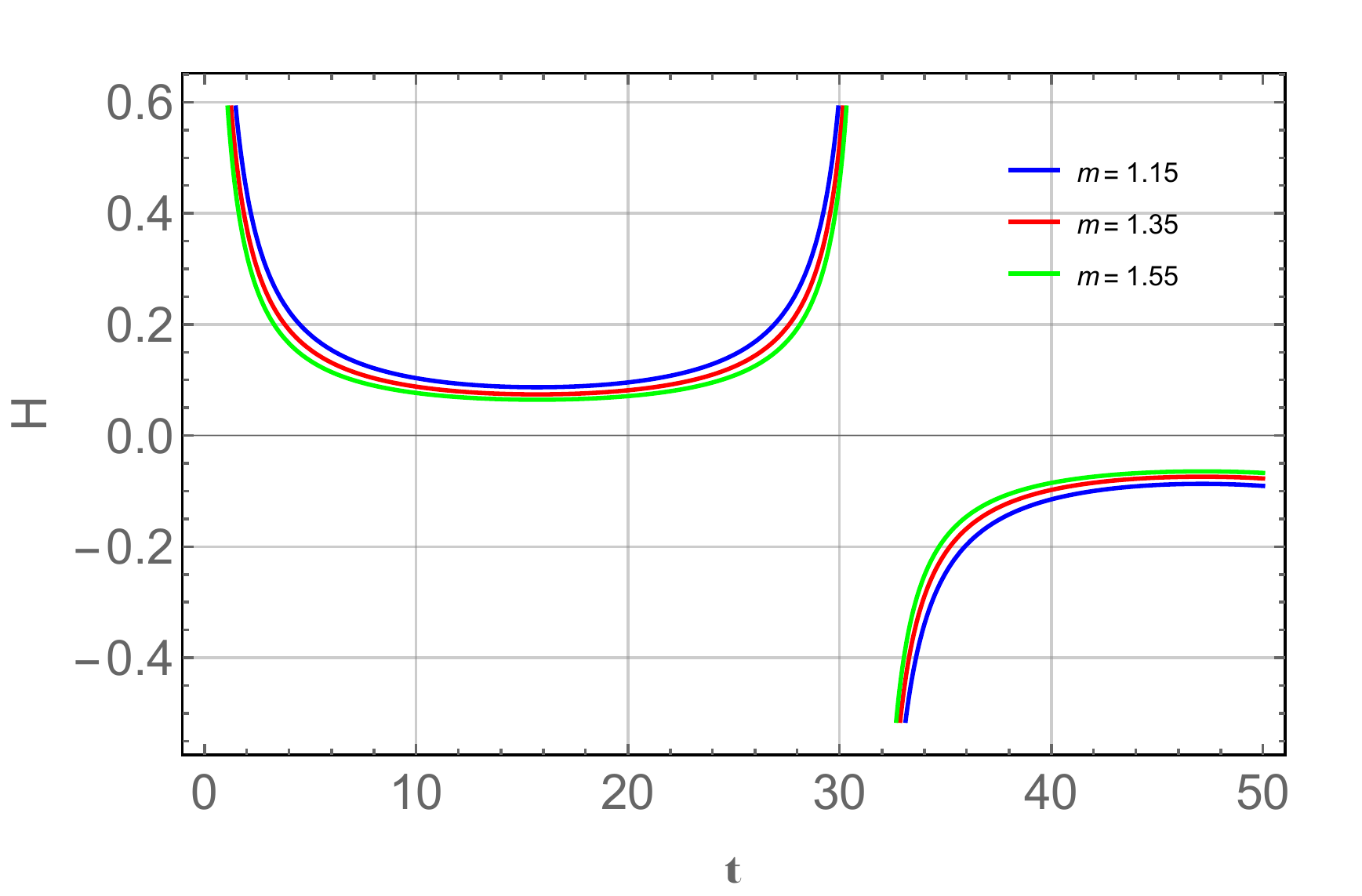}
  \caption{Hubble parameter as a function of cosmic time for $k=0.1$ and three representative values of $m$.}\label{fig4}
\end{figure}


The scale factor and the Hubble parameter derived from the PVDP are shown  in Fig. 3 and Fig. 4 respectively for some specific time frame. Within the time frame, the scale factor increases with cosmic time whereas the Hubble parameter decreases with time. However, the evolutionary behaviour of the scale factor is governed by a $tan$ function and that of the Hubble parameter is governed by a $sine$ function and therefore both can either be positive or negative at some epoch. 

\section{Dynamical properties of the model}\label{IV}

The assumed dynamics of the universe with a PVDP helps us to study the other dynamical properties of the model. The energy density and pressure are obtained from Eqs. \eqref{eq:5}-\eqref{eq:6} as

\begin{eqnarray}
\rho &=&\frac{(3+6\lambda)H^2-2\lambda \dot{H}}{(1+3\lambda)^2-\lambda^2},\label{eq:12}\\ 
p &=& \frac{-(3+6\lambda)H^2-2(1+3\lambda) \dot{H}}{(1+3\lambda)^2-\lambda^2}.\label{eq:13}
\end{eqnarray}

For a PVDP as defined Eq. \eqref{eq:7}, the above expressions reduce to 
\begin{eqnarray}
\rho &=&\left[\frac{2\lambda m\cos kt+3(2\lambda+1)}{(3\lambda+1)^2-\lambda^2}\right]\frac{k^2}{m^2\sin^2 kt},\label{eq:14}\\
p &=& \left[\frac{2(3\lambda+1) m\cos kt-3(2\lambda+1)}{(3\lambda+1)^2-\lambda^2}\right]\frac{k^2}{m^2\sin^2 kt}.\label{eq:15}
\end{eqnarray}
\begin{figure}
\centering
\includegraphics[width=75mm]{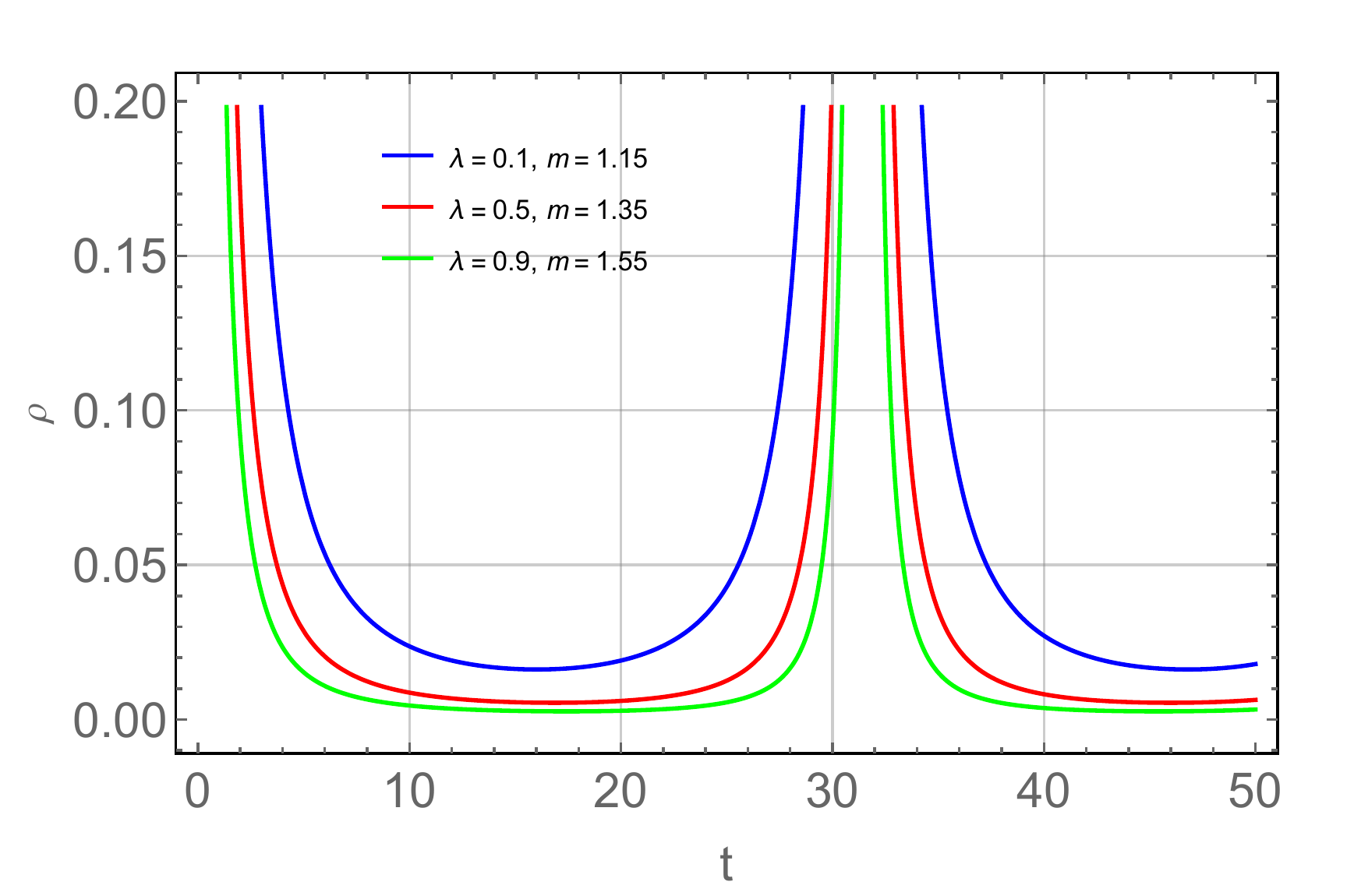}
  \caption{Time variation of energy density.}\label{fig5}
\end{figure}
\begin{figure}
\centering
  \includegraphics[width=75mm]{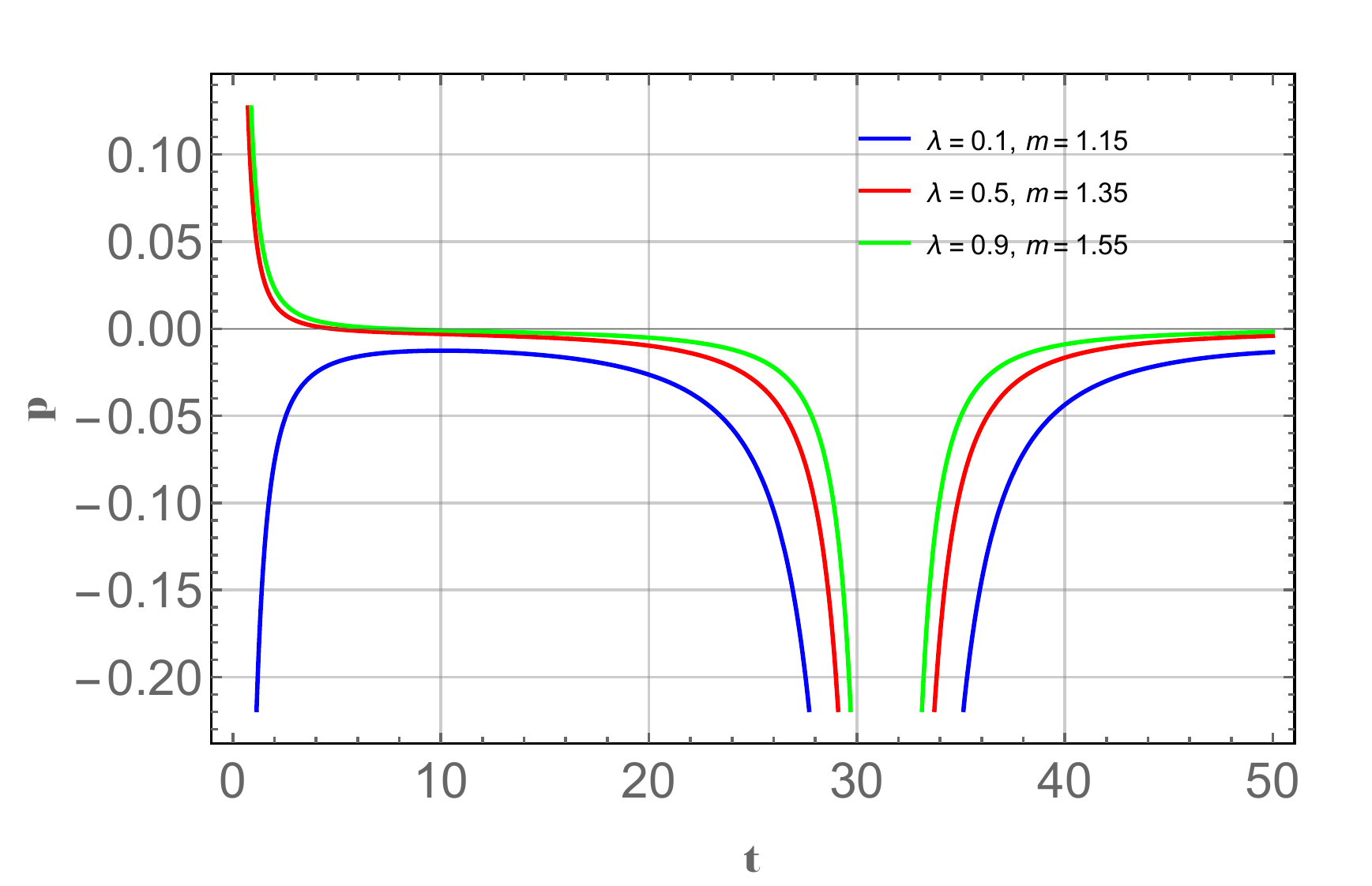}
  \caption{Time variation of pressure.}\label{fig6}
\end{figure}

Equation \eqref{eq:14} sets up a condition for the parameters $\lambda$ and $m$ to get a positive energy density, 
\begin{equation}
6\lambda+3 > 2\lambda m.\label{eq:16}
\end{equation}
A signature flipping behaviour of the deceleration parameter fixes  $m$ to be greater than 1 (refer Fig.\ref{fig2}). We have constrained $m$ from the cosmic transit redshift $z_{da}$ to be $1.55$. From Eq. \eqref{eq:16}, it is certain that, this value of $m$ allows any positive values for $\lambda$. In view of this, one may take $\lambda$ as a free parameter with positive values only. In the present work, we have considered three moderate values, $\lambda = 0.1,0.5$ and $0.9$ for numerical calculation of the dynamical parameters.  In Fig. \ref{fig5}, the periodic variation of the energy density is shown for these values of the parameter $\lambda$. It is evident that, energy density has periodic singularities at the cosmic times $t=\frac{n\pi}{k}$, $n=0,1,2,3,\cdots$ is an integer. The periodic variation clearly depends on the choice of $k$. Since we have taken $k=0.1$, the cosmic singularity occurs corresponding to the time period, $t=0, 31.4, 62.8, \cdots$. The interesting feature is that, in a given cosmic cycle, it starts from a very large value at an initial time ($t\rightarrow 0$) and decreases to a minimum, $\rho_{min}$, and then again increases with the growth of time. The minimum in energy density occurs at a time given by $t=\frac{(n+1)\pi}{2k}$. The evolutionary trend of the energy density is not changed by a variation of $\lambda$,  rather an increase in $\lambda$ simply decreases the value of $\rho$ at a given time. In other words, with an increase in $\lambda$, there occurs a decrement in $\rho_{min}$.

The evolutionary behaviour of pressure is shown in Fig. \ref{fig6}. Pressure is also found to have a periodic variation with singularities at $t=\frac{n\pi}{k}$ or at  $t=0,31.4, 62.8, \cdots$. Within a given cycle, pressure decreases from large positive values at the beginning to large negative values and then reverses the trend. However, pressure is a negative quantity at the present epoch in a given cycle. The choice of the parameter $\lambda$ has some effects on the evolutionary trend. In a given cycle, in general, lower value of $\lambda$ results in a pressure curve that lies to the left side in the figure. The crossing over time from positive domain to negative domain is decided by the value of $\lambda$. 

The equation of state parameter (EOS) $\omega=\frac{p}{\rho}$ can be obtained in a straightforward manner from Eqs. \eqref{eq:14}  and \eqref{eq:15} as
\begin{equation}
\omega=\frac{2 (3 \lambda +1) m \cos k t-3(2\lambda +1)}{2 \lambda m \cos k t+3(2 \lambda +1)}.\label{eq:17}
\end{equation}

\begin{figure}
\centering
\includegraphics[width=75mm]{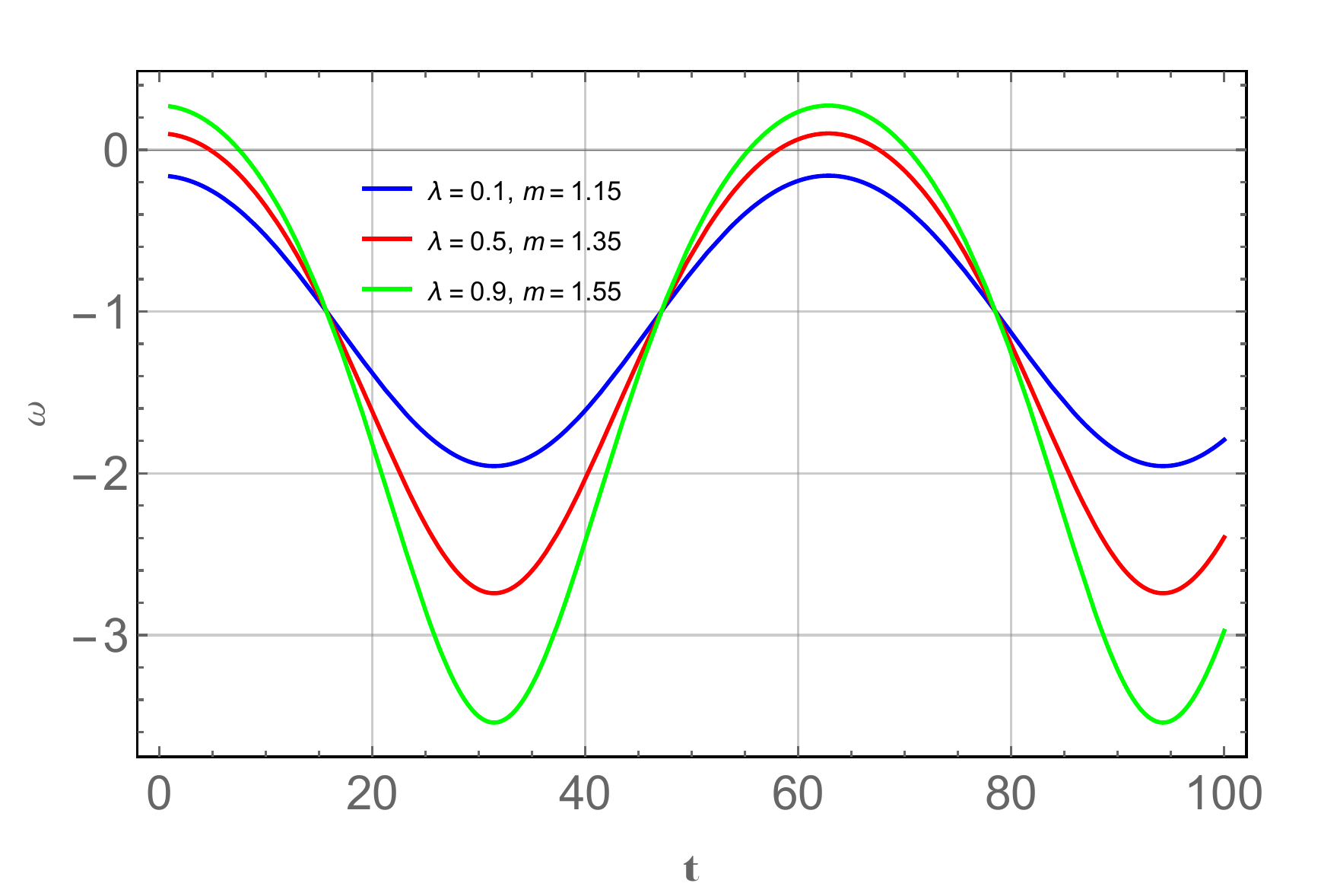}
  \caption{Equation of state parameter as function of cosmic time.}\label{fig7}
\end{figure}

\begin{figure}
\centering 
  \includegraphics[width=75mm]{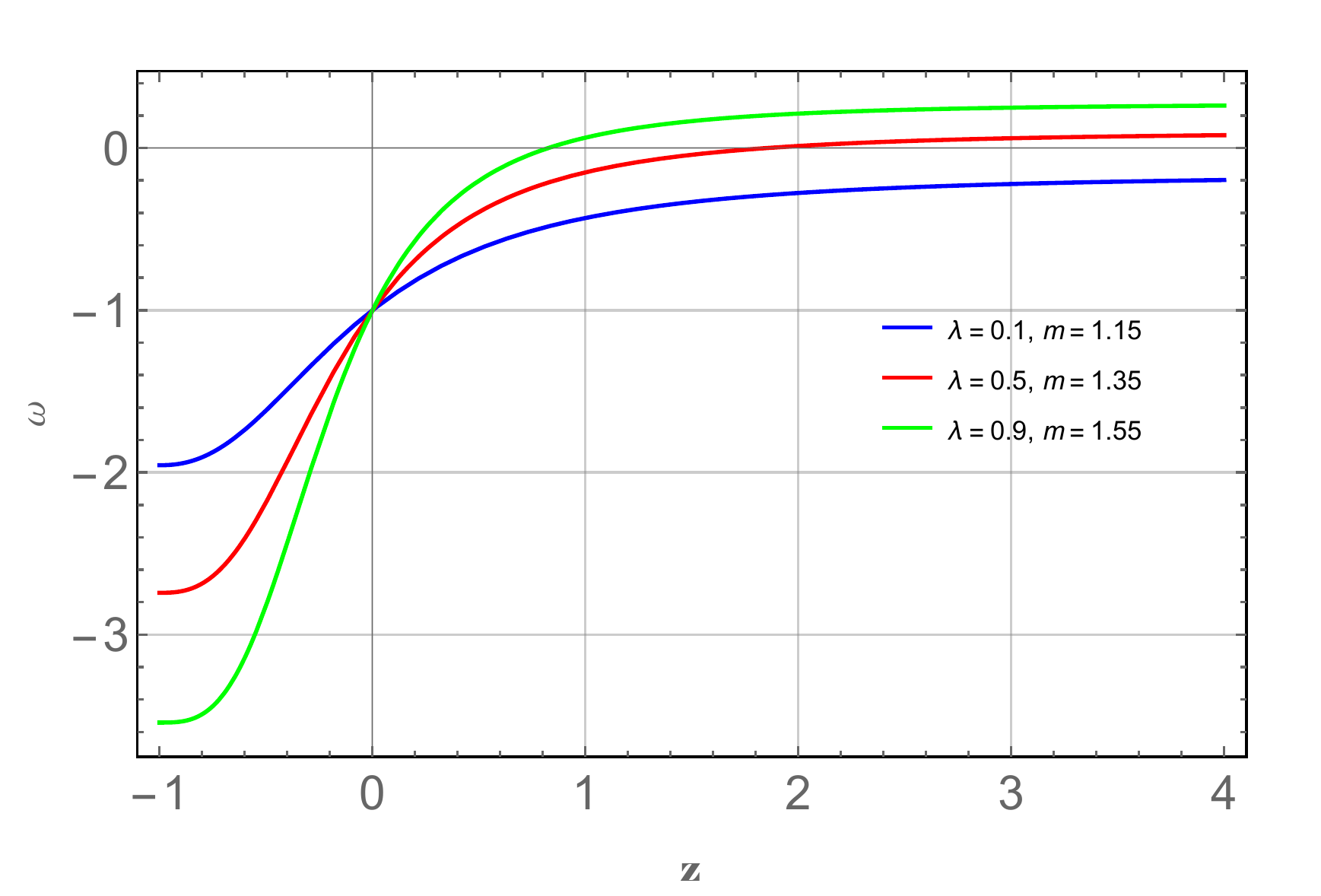}
  \caption{Equation of state parameter as function of redshift.}\label{fig8}
\end{figure}

In Fig. \ref{fig7}, we have plotted the evolution of EOS parameter, $\omega$, as a function of cosmic time. The EOS parameter exhibits an oscillatory behaviour. In the first half of the cosmic cycle, $\omega$ decreases from a positive value close to $\frac{1}{3}$ to negative values after crossing the phantom divide at $\omega =-1$. After attending a minimum it again increases to positive value at the end of the cycle. One interesting feature of the equation of state parameter is that, unlike the energy density and pressure, it does not acquire any singular values during the cosmic cycle. This fact is due to the cancellation of the $\frac{1}{\sin^2 kt}$ factor from the pressure by the same factor of energy density and depends only the value of the PVDP. Since the PVDP does not a have singularity, the same thing also occurs in the EOS parameter. The oscillatory behaviour comes only from the $\cos kt$ factor appearing both in the numerator and denominator of $\omega$. The evolutionary trend is affected by the choice of $\lambda$. Curves of $\omega$ with low values of $\lambda$ remain on top before the phantom divide whereas after the phantom divide, they remain in the bottom of all the curves. At an equivalent present epoch $\left(t=\frac{8n+1}{k}\frac{\pi}{2}\right)$ in any given cycle, the EOS parameter remains within the quintessence region with a value close to $-1$. One can decipher the evolution of the EOS parameter, in the purview of usual Friedman model, as an evolution from a radiation dominated phase ($\omega = \frac{1}{3}$) to a matter dominated phase ($\omega =0 $) and then to a dark energy driven accelerated phase ($\omega < -\frac{1}{3}$). At the present epoch, our model predicts an EOS that behaves more like a cosmological constant and the model is somewhat close to that of $\Lambda$CDM model. This aspect of the EOS parameter is clearly visible in Fig. 8, where we have shown the evolution of $\omega$ as a function of redshift. It is clear from Fig. \ref{fig8} that, at $z=0$, $\omega =-1$. It is interesting to note that, the overlapping of the present model with $\Lambda$CDM at equivalent present epochs is independent of the choice of the parameter $\lambda$. In this model from Fig. \ref{fig8}, it can be observed that the evolution of $\omega$ as a function of redshift shows a transitional behavior from  $\omega <-1$ at low redshift to $\omega>-1$ at higher redshift. This transitional behavior of $\omega$ of this model fits with SNL3 data \cite{Feng2005}. In particular, the use of SNL3 data suggests that BAO data is also partly responsible for this. Some recent reconstruction of the EOS from different observational data sets including the high redshift Lyman-$\alpha$ forest (Ly$\alpha$FB) measurement favours a non-constant dynamical dark energy. In these reconstructed models, the EOS evolves with time and crosses the phantom divide \cite{Zhao12, Zhao17}. The behavior of EoS $\omega$ in this model is consistent with quintom model which allows  $\omega$ to cross $-1$. As inferred in Ref.\cite{Zhao17}, the departure from $\omega=-1$ is more evident in the reconstruction history of the dynamical dark energy with more recent data sets including the Ly$\alpha$FB measurement.


For the present model with a PVDP, we obtain, the density parameter $\Omega = \frac{\rho}{3H^2}$ as 
\begin{equation}
\Omega =\frac{1}{3}\left[\frac{2\lambda m\cos kt+3(2\lambda+1)}{(3\lambda+1)^2-\lambda^2}\right],\label{eq:18}
\end{equation}

which can be expressed as a function of redshift as

\begin{equation}
\Omega =\frac{1}{3}\left[\frac{2\lambda m\cos \left(2 \tan ^{-1}\frac{1}{(z+1)^m}\right)+3(2\lambda+1)}{(3\lambda+1)^2-\lambda^2}\right],\label{eq:19}
\end{equation}

\begin{figure}[t]
\centering
\includegraphics[width=75mm]{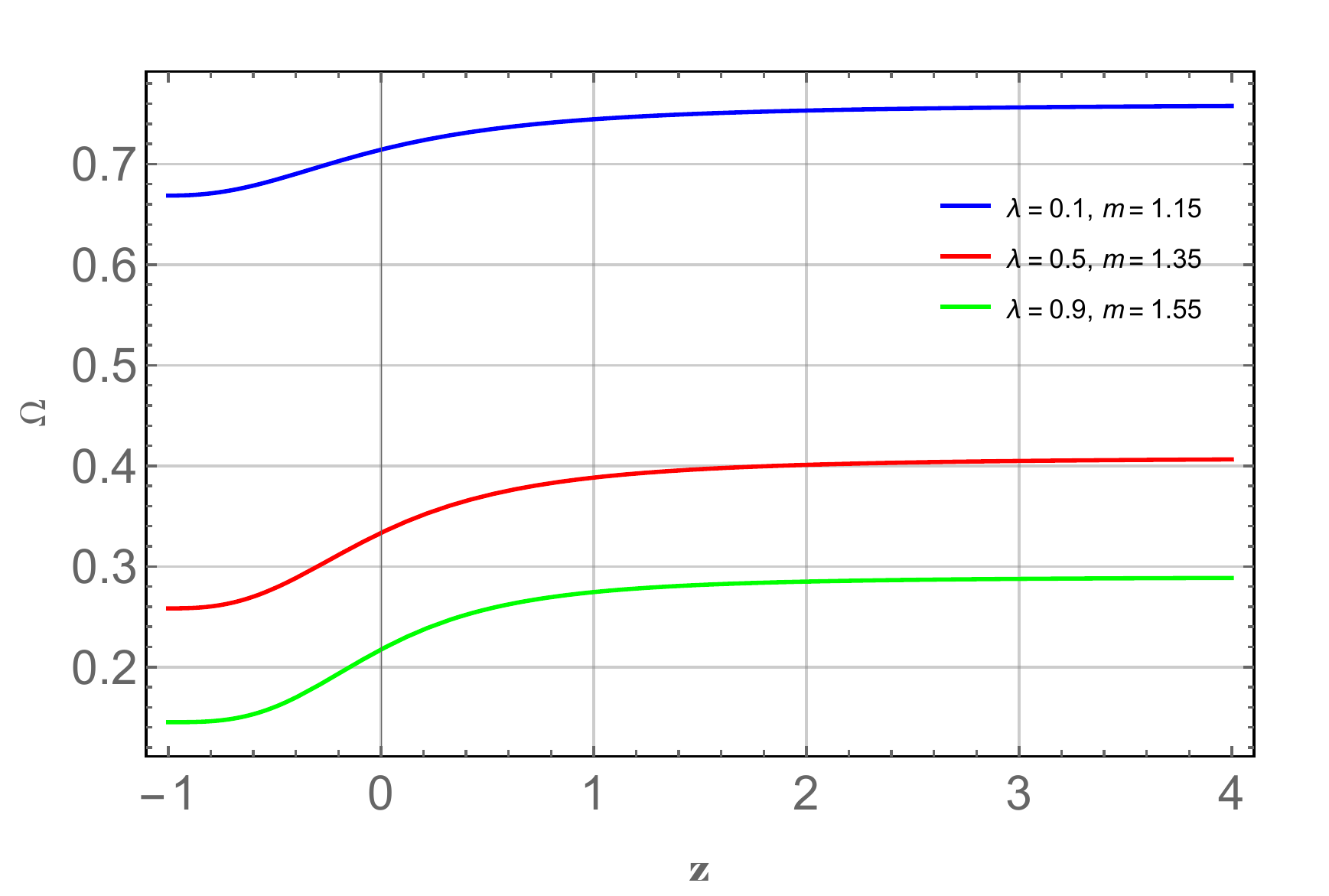}
\caption{Evolution of Density parameter.}\label{fig9}
\end{figure}

The density parameter as a function of redshift is plotted in Fig. \ref{fig9} for three representative values of $\lambda$. $\Omega$ remains almost unaltered in the range of redshift greater than 1 for all the three values of $\lambda$ considered in the work. However, with the cosmic evolution,  $\Omega$ decreases with cosmic time after $z=1$. The density parameter, at a given redshift, is observed to have lower value for higher values of $\lambda$.

%

\section{Violation of Energy-Momentum Conservation}\label{V}

\begin{figure}[h!]
\centering
\includegraphics[width=75mm]{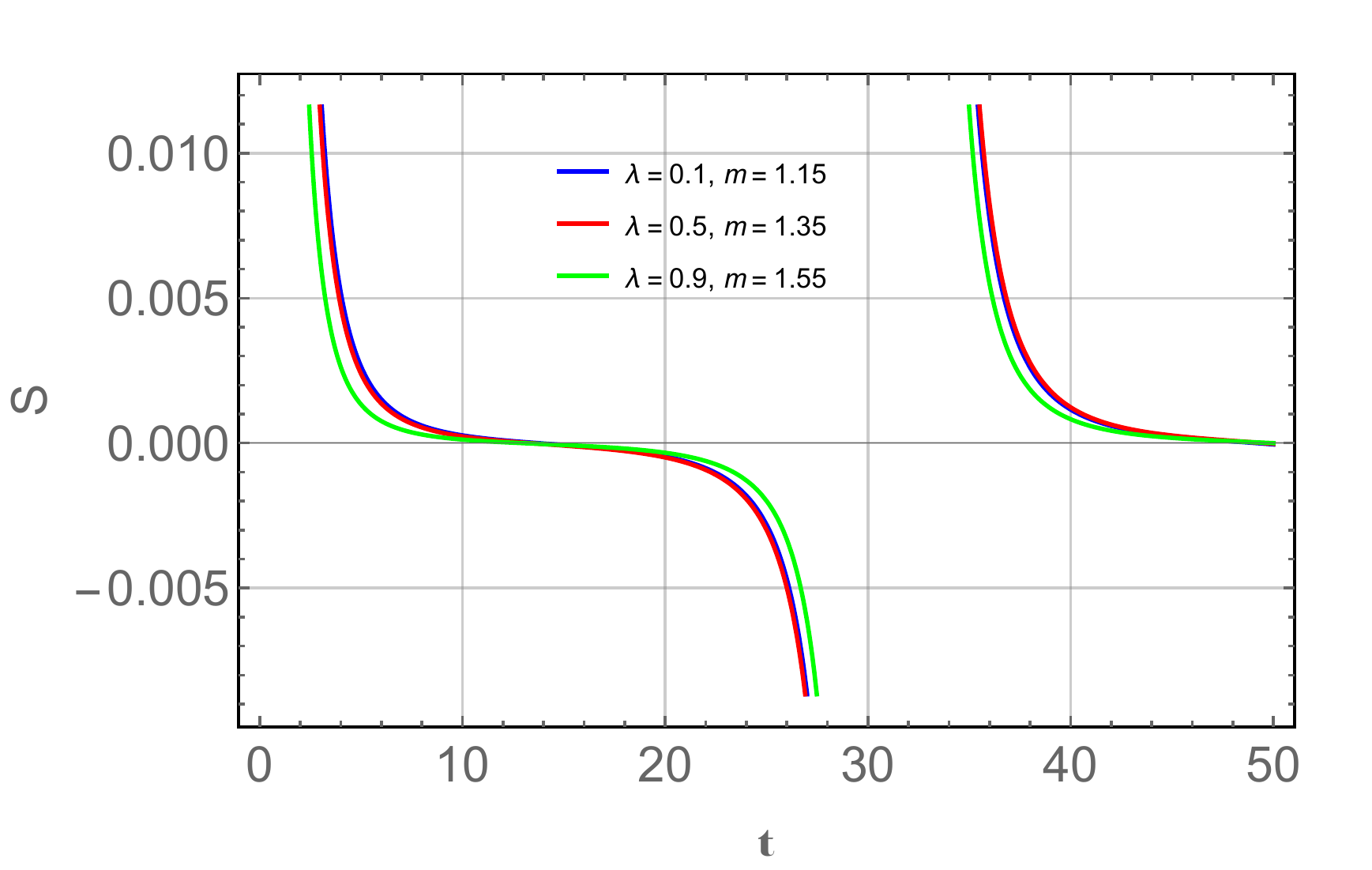}
\caption{Energy-momentum non conservation.}\label{fig11a}
\end{figure}

Friedman models in GR ensure the energy conservation through the continuity equation
\begin{equation}
\dot{\rho}+3H(\rho+p)=0, 
\end{equation}
which implies $d(\rho V)=-pdV$. Here $V=a^3$, the volume scale factor of universe and the quantity $\rho V$ gives an account of the total energy. As the universe expands the amount of dark energy in an expanding volume increases in proportion to the volume. If that space time is standing completely still, the total energy is constant; if it's evolving, the energy changes in a completely unambiguous way. It decreases or increases in time. However in modified gravity theories, one may get a different picture. Taking a covariant derivative of Eq. \eqref{eq:2}, one can obtain \cite{Harko14, alva13, bar14, das17} 

\begin{multline}
\nabla^{i}T_{ij}=\frac{\mathcal{F(R,T)}}{\kappa-\mathcal{F(R,T)}}[(T_{ij}+\Theta_{ij})\nabla^{i} ln \mathcal{F(R,T)} \\ + \nabla^{i}\Theta_{ij} -\frac{1}{2}g_{ij}\nabla^{i}T]. \label{eq:21}
\end{multline}
With the substitution of $f(R,T)=R+2\lambda T$ and $\kappa =1$, Eq. \eqref{eq:21} reduces to
\begin{equation}
\nabla^{i}T_{ij}=-\frac{2\lambda}{1+2\lambda} \left[\nabla^{i}(pg_{ij})+\frac{1}{2}g_{ij} \nabla^{i}T\right].\label{eq:22}
\end{equation}
One should note here that, for $\lambda =0$, one would get $\nabla^{i}T_{ij} =0$. However for $\lambda \neq 0$, the conservation of energy-momentum is violated. Recently some researchers have investigated the consequence of the violation of energy-momentum conservation (i.e. $\dot{\rho}+3H(\rho+p) \neq0 $) in modified gravity theories. The non-conservation of energy-momentum may arise due to non unitary modifications of quantum mechanics and in phenomenological models motivated by quantum gravity
theories with spacetime discreteness at the Planck scale \cite{Josset17}. In the context of unimodular gravity, Josset et al. have shown that a non-conservation of energy-momentum leads to an effective cosmological constant which increases or decreases with the creation or annihilation of energy during the cosmic expansion and can be reduced to a constant when matter density diminishes \cite{Josset17}. Shabani and Ziaie \cite{Shabani17} have studied the consequence of the non-conservation of energy-momentum in some classes of $f(R,T)$ gravity with pressure-less cosmic fluid and showed a violation of energy-momentum conservation in modified theories of gravity can provide accelerated expansion. Also, the non-conservation of the energy-momentum tensor implies in non-geodesic
motions for test particles in gravitational fields as it was deeply investigated \cite{Baffou17}. In \cite{Shabani18,Moraes18} the authors have constructed a formalism in which an effective fluid is conserved in $f(R,T)$ gravity, rather than the usual energy-momentum tensor non-conservation. We have investigated the non-conservation of energy-momentum for the class of models in $f(R,T)$ gravity with the suggested PVDP. We quantify the violation of energy-momentum conservation through a deviation factor $S$, defined as

\begin{equation}
S=\dot{\rho}+3H(\rho+p).\label{eq:23}
\end{equation}
In case of the model satisfying energy-momentum conservation, we have $S=0$, otherwise, we get a non zero value for this quantity. $S$ can be positive or negative depending on whether the energy flows away from or into the matter field. In Fig. \ref{fig11a}, we have shown the non-conservation of energy-momentum for a periodic cosmic cycle. It is clear that, except for a very limited period, the conservation is violated along with the cosmic evolution. However, the nature of energy flow changes periodically. This behaviour is obtained for all the values of $\lambda$ considered in the work. At an equivalent present epoch in a given cosmic cycle, at least within the purview of the present model, there is a signal of non-conservation.

\section{Energy Conditions}\label{VI}

\begin{figure}[h!]
\centering
\includegraphics[width=75mm]{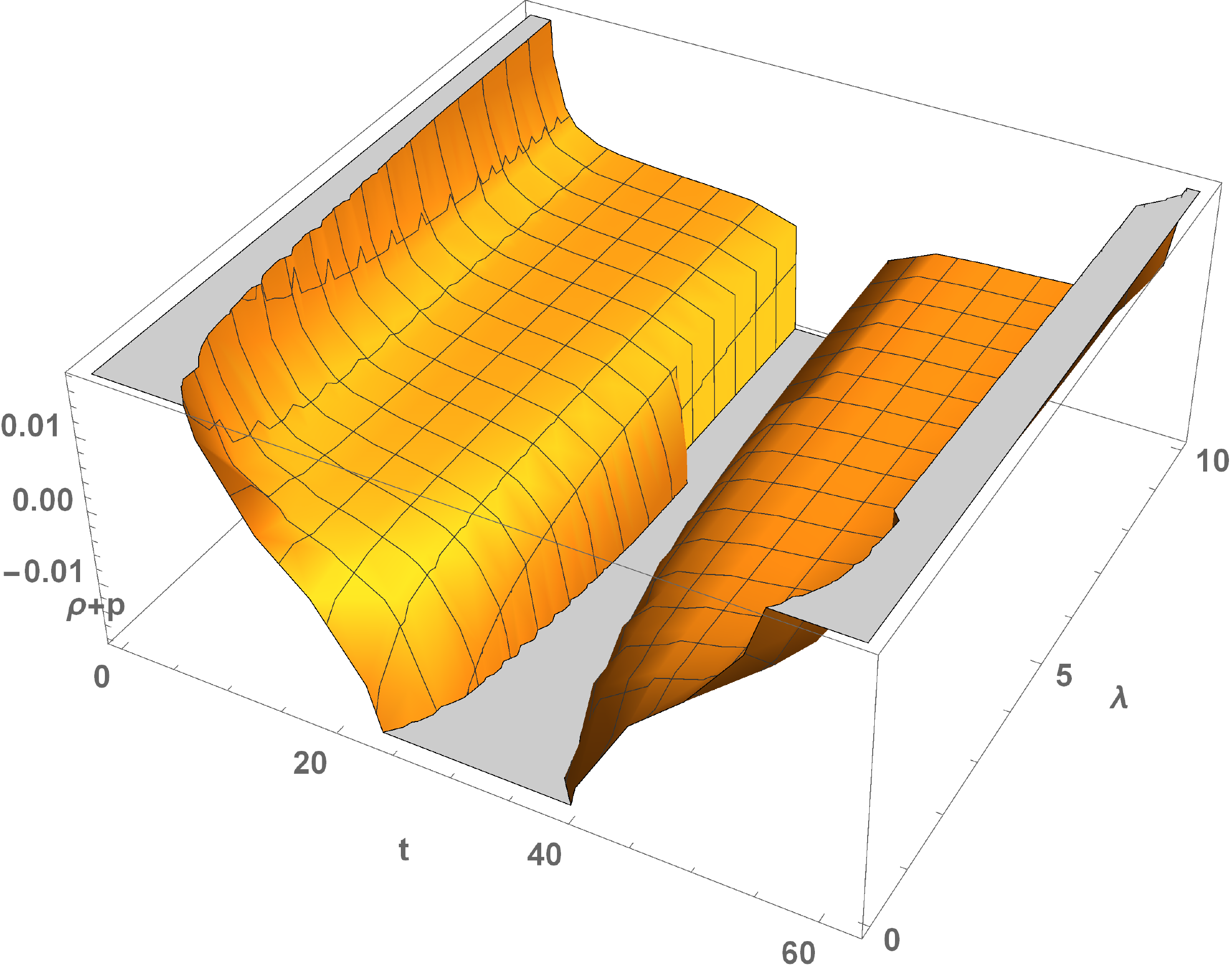}
  \caption{ Violation of NEC ($\rho+p \geq 0$) versus $\lambda$ and $t$.}\label{fig11}
\end{figure}

\begin{figure}[h!]
\centering
  \includegraphics[width=75mm]{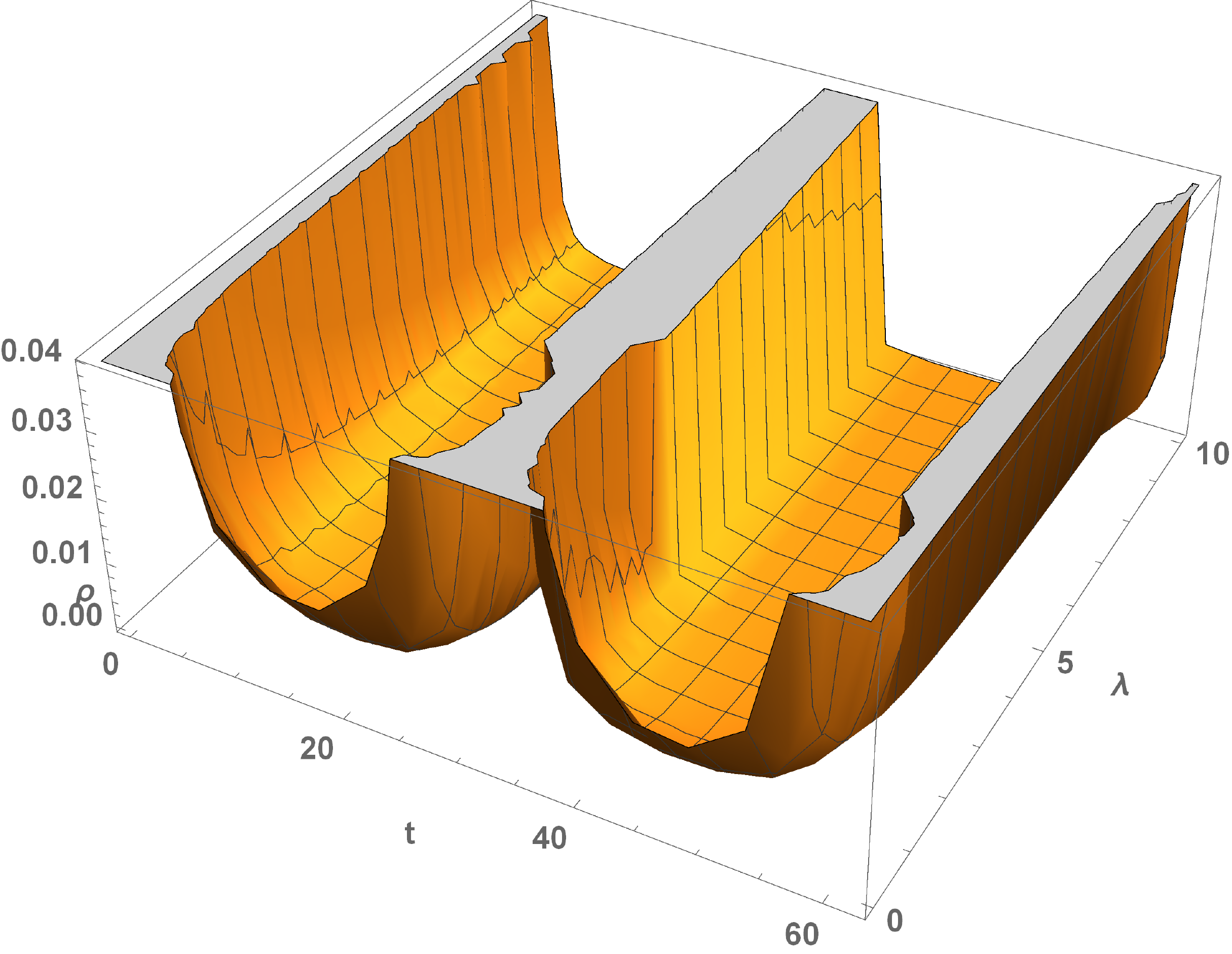}
  \caption{$\rho \geq 0$ versus $\lambda$ and $t$.}\label{fig12}
\end{figure}

\begin{figure}[h!]
\centering
\includegraphics[width=75mm]{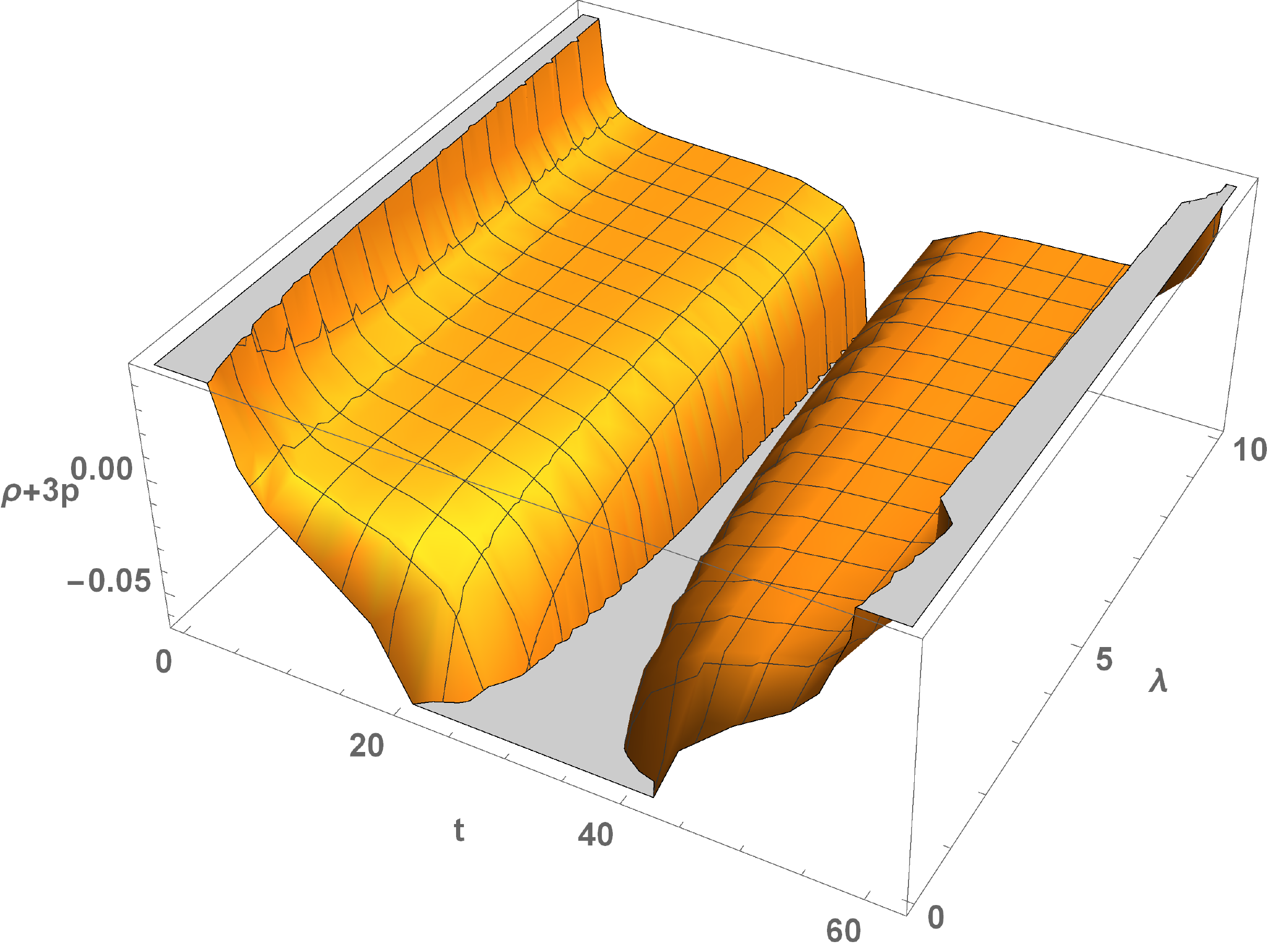}
  \caption{Violation of SEC ($\rho+3p \geq 0$) versus $\lambda$ and $t$.}\label{fig13}
\end{figure}

\begin{figure}[h!]
\centering
  \includegraphics[width=75mm]{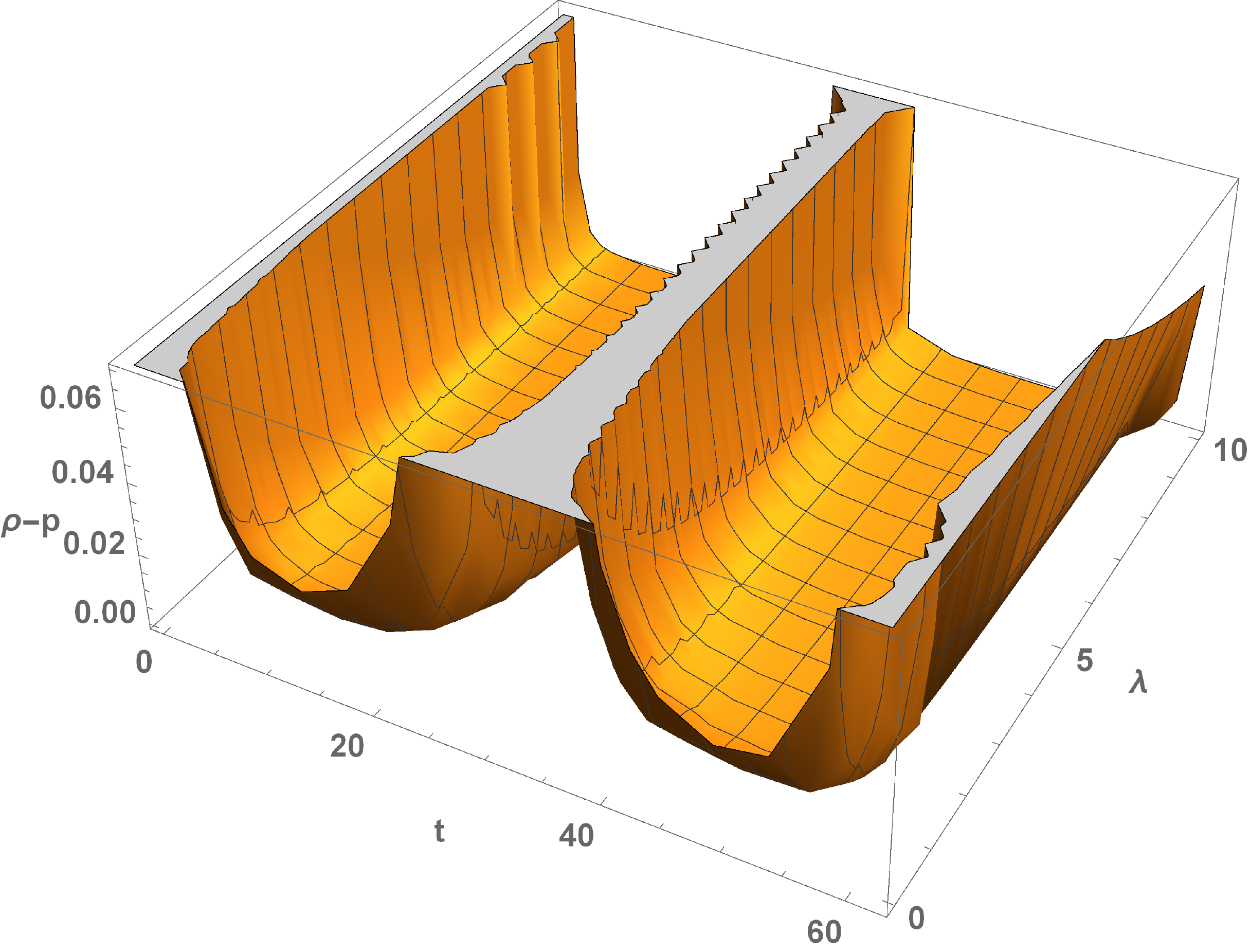}
  \caption{DEC, $\rho \geq \vert p\vert $ versus $\lambda$ and $t$.}\label{fig14}
\end{figure}

The energy conditions (ECs) of GR are a variety of different ways to show positive energy density more precisely. The ECs
take the form of various linear combinations of the stress-energy tensor components (at any specified point in spacetime) should be positive, or at least non-negative \cite{Visser/2000}. The ECs are the fundamental tools for the study of wormholes (WHs) and black holes in different physical scenario. Particularly ECs are the outcomes of Raychaudhuri's equation for the
expansion nature \cite{Carroll/2004}. The study of singularities in the spacetime was based on ECs. The main ECs in GR such as null, weak, strong, and dominant (NEC, WEC, SEC, and DEC) respectively for the energy-momentum tensor are expressed as
\begin{eqnarray}
\text{NEC} \Leftrightarrow \rho+p \geq 0,\\ 
\text{WEC} \Leftrightarrow \text{NEC} \ \ \text{and} \ \ \rho \geq 0,\\
\text{SEC} \Leftrightarrow \rho+3p \geq 0,\\
\text{DEC} \Leftrightarrow \rho \geq \vert p\vert .
\end{eqnarray}
Particularly, Alvarenga et al. \cite{Alvarenga/2012} and Sharif et al. \cite{Sharif/2013} have analyzed the ECs in $f(R,T)$ gravity. The above energy conditions are plotted in Figures \ref{fig11}-\ref{fig14} with $m=1.55$, $k=0.1$ and varying $\lambda$.


One can observe from the above Figures \ref{fig11}-\ref{fig14}, all the ECs are behaved periodic for fixed values of $m$ and $k$ with accepted range for $\lambda$ in this model. The free parameters are considered based on the positivity of energy density as shown in Figure-\ref{fig12}. In the present model DEC is satisfied and all other ECs are violated.

\section{Stability Analysis}\label{VIIa}
In this section we wish to analyse the stability of our model under linear homogeneous perturbations in the FRW background. We consider linear perturbations for the Hubble parameter and the energy density as \cite{Sharif2013}
\begin{eqnarray}
H(t)&=& H_b(t)\left(1+\delta(t)\right),\label{eq:28}\\
\rho(t)&=& \rho_b\left(1+\delta_m(t)\right),\label{eq:29}
\end{eqnarray}
where $\delta (t)$ and $\delta_m(t)$ are the perturbation parameters. In the above, we have assumed a general solution $H(t)=H_b(t)$ which satisfies the background FRW equations. The matter energy density can be expressed in terms of $H_b$ as 
\begin{equation}
\rho_b = \frac{(3+6\lambda)H_b^2-2\lambda \dot{H}_b}{(1+3\lambda)^2-\lambda^2}.
\end{equation}

The Friedman equation and the trace equation for the modified gravity model with a functional $f(R,T)=R+2\lambda T$ can be obtained as
 
\begin{eqnarray}
\mathbf{\Theta}^2 &=& 3[\rho+2\lambda (\rho+p)+f(R,T)],\label{eq:30}\\
R &=& -(\rho-3p)-2\lambda (\rho+p)-4f(R,T).\label{eq:31}
\end{eqnarray}
Here, $\mathbf{\Theta}=3H$ is the expansion scalar. For a standard matter field, we can have the first order perturbation equation

\begin{equation}
\dot{\delta}_m(t)+3H_b(t)\delta(t)=0.\label{eq:32}
\end{equation}

Using Eqs.\eqref{eq:28}- \eqref{eq:30}, one can obtain
\begin{equation}
(1+3\lambda)T\delta_m(t)=6H_b^2\delta(t).\label{eq:33}
\end{equation}
The first order matter perturbation equation can be obtained by the elimination of $\delta(t)$ from Eqs. \eqref{eq:32} and \eqref{eq:33} as
\begin{equation}
\dot{\delta}_m(t)+\frac{T}{2H_b}\left(1+3\lambda\right)\delta_m(t)=0.\label{eq:34}
\end{equation}
Integration of Eq.\eqref{eq:34} leads to 
\begin{equation}
\delta_m(t)= C ~exp\left[-\left(\frac{1+3\lambda}{2}\right)\int \frac{T}{H_b}~dt \right],
\end{equation}
where $C$ is a non zero positive constant. Consequently, the evolution of the perturbation $\delta(t)$ becomes
\begin{equation}
\delta(t)=\frac{(1+3\lambda)CT}{6H_b^2}~exp\left[-\left(\frac{1+3\lambda}{2}\right)\int \frac{T}{H_b}~dt \right].
\end{equation}

Since 
\begin{equation}
\frac{T}{H_b}=\frac{-k}{[(1+3\lambda)^2-\lambda^2]}\left[\frac{(16\lambda+6)\cos kt}{\sin kt}+\frac{6(2\lambda+1)}{m~\sin kt}\right],
\end{equation}
the factor $\int \frac{T}{H_b}~dt$ is evaluated as

\begin{equation}
\int \frac{T}{H_b}dt= -\frac{\splitfrac{(6 \lambda +(8 \lambda +3) m+3) \log (1-\cos (k t))}{+(-6 \lambda +(8 \lambda +3) m-3) \log (\cos (k t)+1)}}{\left(8 \lambda ^2+6 \lambda +1\right) m}.
\end{equation}

The growth and decay of the perturbation depend on the factors $k$ and $\lambda$ periodically. We found that the considered values for $k$ and $\lambda$ in the physical parameters $\rho$, $p$ and $\omega$ are compatible with the decay of perturbation.

\section{Conclusion}\label{VII}

In this work, we have studied the background cosmology of an isotropic flat universe in the framework of $f(R, T)$ gravity.  According to the observations and associated analysis, the universe undergoes an accelerated expansion in the present epoch. The universe might have transitioned from a decelerated phase at some past epoch to an accelerated phase. This behaviour clearly hints for a time varying deceleration parameter which should evolve from a positive value in past to negative values at late phase of cosmic time. In other words, the evolving deceleration parameter displays a signature flipping behaviour. Keeping in view the signature flipping nature of the deceleration parameter, in the present work, we assume a periodically varying deceleration parameter to reconstruct the cosmic history. The PVDP has two adjustable parameters one of which can be constrained from the cosmic transit behaviour. The assumed deceleration parameter oscillates in between two limits usually set by the transit redshift. Consequently, the universe in this model, starts with a decelerating phase and evolves into a phase of super-exponential expansion with a periodic repetition of the phenomenon. 

The PVDP obviously generates periodically varying dynamical properties of the universe. The energy density and pressure 
vary cyclically within a given cosmic period decided by the cosmic frequency parameter of the model. At some finite time, the magnitude of these physical parameters become infinitely large. This behaviour leads to a future type I singularity as classified by Nojiri et al \cite{Noj2005}. There appears to have a big Rip at certain finite time during the cosmic repetition of the phenomenon since $a\rightarrow \infty, \rho \rightarrow \infty$ and $|p|\rightarrow \infty$. Since the parameters repeats their behaviour after a time period $t=\frac{n\pi}{k}$, the big Rip also occurs  periodically after a time gap of $t=\frac{n\pi}{k}$.

The equation of state parameter for the PVDP has a cyclic behaviour that repeats with time. The EOS parameter evolves from a radiation dominated phase ($\omega = \frac{1}{3}$) to a matter dominated phase ($\omega =0 $) and then to a dark energy driven accelerated phase ($\omega < -\frac{1}{3}$). It may cross the phantom divide $\omega=-1$ for some cosmic time range. How far the well of the EOS parameter will go beyond the phantom divide is decided by the coupling constant $\lambda$. The EOS acquires a deeper well for a larger value of coupling constant.  At the present epoch, our model predicts an EOS that behaves more like a cosmological constant.

In the present work, we have investigated an important aspect of $f(R,T)$ gravity theory, the violation of energy-momentum conservation. It is well known that, in modified theories like $f(R,T)$ gravity, the energy-momentum conservation is violated and this non-conservation can lead to a sort of accelerated expansion of the universe. $f(R,T)$ theories of gravity, as originally proposed, predict a non-conservation of the usual energy-momentum tensor of matter \cite{Harko2011}. As a result, although the form of the field equations remain same but now the test particles move in a geodesics and the choice of the Lagrangian function is not totally arbitrary. We have shown that, a PVDP leads to a kind of universe model, where within a given cycle, the energy momentum conservation is continuously violated except for a small period of cosmic time. As pointed out by Josset and Perez \cite{Josset17}, modified gravity models can explain the accelerated expansion at the cost of energy momentum non conservation. 

We have analyzed the energy conditions for the $f(R,T)$ modified gravity theory in a
general way. For the perfect fluid source of matter in Sect. \ref{VI} we have shown the violation/validity of the
energy conditions both analytically as well as graphically. The energy conditions in modified theories of gravity have a well
defined physical motivation, i.e., Raychaudhuri's equation along with attractive nature of gravity. For the considered
model of $f(R,T)$ gravity, the NEC and SEC are
derived from the Raychaudhuri equation together with the
condition that gravity is attractive. It is shown that these
conditions behaved periodic from those derived in the context of GR. Finally, we have discussed the stability of the solutions under linear homogeneous perturbations. The stability depends on the values of the parameters $k$ and $\lambda$ as they behaves periodically.

\section*{Acknowledgments}
PKS and PS acknowledges DST, New Delhi, India for providing facilities through DST-FIST lab, Department of Mathematics, where a part of this work was done. The authors also thank the referees for their valuable suggestions, which improved the presentation of the present results.

\end{document}